# The Impact of Generative AI on the Future of Visual Content Marketing

Shiva Mayahi

Islamic Azad University, Karaj Branch, Iran, shivamaya98@gmail.com

Marko Vidrih

University of Ljubljana, Slovenia, marko@elvivia.si

In today's world of marketing, it is necessary to have visually appealing content. Visual material has become an essential area of focus for every company as a result of the widespread availability of gadgets for mass communication and extended visual advancements. Similarly, artificial intelligence is also gaining ground and it is proving to be the most revolutionary technological advancement thus far. The integration of visual content with artificial intelligence is the key to acquiring and retaining loyal customers; its absence from the overarching marketing strategy of any production raises a red flag that could ultimately result in a smaller market share for that company.

CCS CONCEPTS • **Computing methodologies ~ Artificial intelligence** • Human-centered computing ~ Human computer interaction (HCI) • **Applied computing ~ Arts and humanities ~ Media arts** • Theory of computation ~ Logic ~ Automated reasoning

**Additional Keywords and Phrases:** Artificial Intelligence, Computational Creativity, Computational Art, AI Art, Generative Media, Generative Design, Generative Art, Computer Art, Creative AI, Visual Communication, Visual Advertising, Visual Content Marketing, AI-generated Designs, Digital Creativity

## 1 INTRODUCTION

In a short time, the world has revolutionized how we look at things. Languages have a fixed quantity of words – and we strive to describe the world around us using the regularly used words and phrases that we know. What we see often cannot be simply explained in words as the mind is primarily visual and sensory-driven. People are innately lured to the highly visual stuff. In addition, the quality of life has improved tremendously, and with it has come to an improvement in aesthetic standards [Xue, 2020]. Brands have realized this and the problem for brands to maximize the amount of quality visual material is one that may be handled using artificial intelligence (AI).

The concept of artificial intelligence, together with the expectations and apprehensions that are linked with its development, is rather prominent in the mind of the average person. Whether we see Judgement Day brought about by Skynet or egalitarian totalitarianism brought about by V.I.K.I and her army of robots, the end effect is the same: the replacement of humans as the preeminent form of life on the planet.

Others would call it a self-fulfilling prophecy, but some people might term it the anxieties of a technophobic mind. If the 2014 research from the University of Reading in the United Kingdom is any indication, we may have already started carrying out the aforementioned prophecy. A computer program allegedly passed the age-old Turing test at the beginning of June 2014, making it the first time in history that this feat had been accomplished. The computer program known as

Eugene Goostman is a name that will forever be ingrained in history. It received praise and criticism from people all over the world, depending on whether they believed it to be the beginning of artificial intelligence or merely a cunning trickster-bot that only demonstrated technical skill.

Vladimir Veselov, who was from Russia, and Eugene Demchenko, who was from Ukraine, came up with the idea for the program in 2001 [University of Reading, 2014]. Since that time, it was modified to mimic the personality and conversational patterns of a boy of 13 years of age, and it was up against four other programs in an effort to emerge victorious. The Turing Exam is widely regarded as the competition that has ever had the most complete test design, and it was staged in London's illustrious Royal Society. In order for a computer program to pass the Turing Test, it must have the ability to fool a human being into believing that the entity with which they are having a conversation is another human being at least thirty percent of the time. This condition is both straightforward and challenging.

As a result of the experiment in London, Eugene received a success rating of 33 percent, making it the first program to successfully complete the Turing Test [University of Reading, 2014]. The test was more difficult overall because it involved simultaneous conversations between human subjects and five different computer programs. There was a total of 300 conversations, with 30 human judges or subjects, and five other computer programs [University of Reading, 2014]. These conversations took place throughout five parallel tests. Only Eugene was successful in persuading 33 percent of the human judges that the youngster in question was actually a human being across all of the tests [University of Reading, 2014]. Built with algorithms that allow "conversational logic" and open-ended themes, Eugene opened up an entirely new realm of intelligent machines capable of tricking human beings.

For many people, the success of this test implied the onset of artificial intelligence, a technological advancement that came with a lot of mixed feelings over the potential threat of gradually but eventually completely eroding human employment. Whereas this still remains a threat, it is also worth considering the positive impact that artificial intelligence will have on certain aspects of our life. One of those is marketing. Today, we have interacted with a fair share of marketing technology that is laced with artificial intelligence, whether knowingly or unknowingly. Most of this has happened on social media through our interactions with various platforms. As it keeps becoming apparent that artificial intelligence is the future of technology, marketers and marketing organizations now need to start thinking about artificial intelligence and how it intertwines with visual marketing, which is the modern-day principle of effective marketing.

## 2   THE USE OF VISUAL CONTENT IN ONLINE MARKETING

There are three primary categories of visual content – illustrations, comics, and videos.  Illustrations can be any message that is communicated through a static design element; this can include paintings, photographs, and memes, which are a more recent development. The picture is the most important component in all of these, and it is accompanied by one or more bits of text. Comics are essentially collections of images and text that are organized in a specific order and have a linear narrative. Infographics are a unique and contemporary type of this graphic that convey vast amounts of information in a graphic manner, often with a visual "narrative" from beginning to end. Lastly, videos refer to moving images in the form of clips, and short films that are able to either tell a story or express a certain message, and ideally both. Important subcategories include vines, which are looping films that are no longer than six seconds in length, and gifs, which are looping slideshows of images that are compressed into a single file.

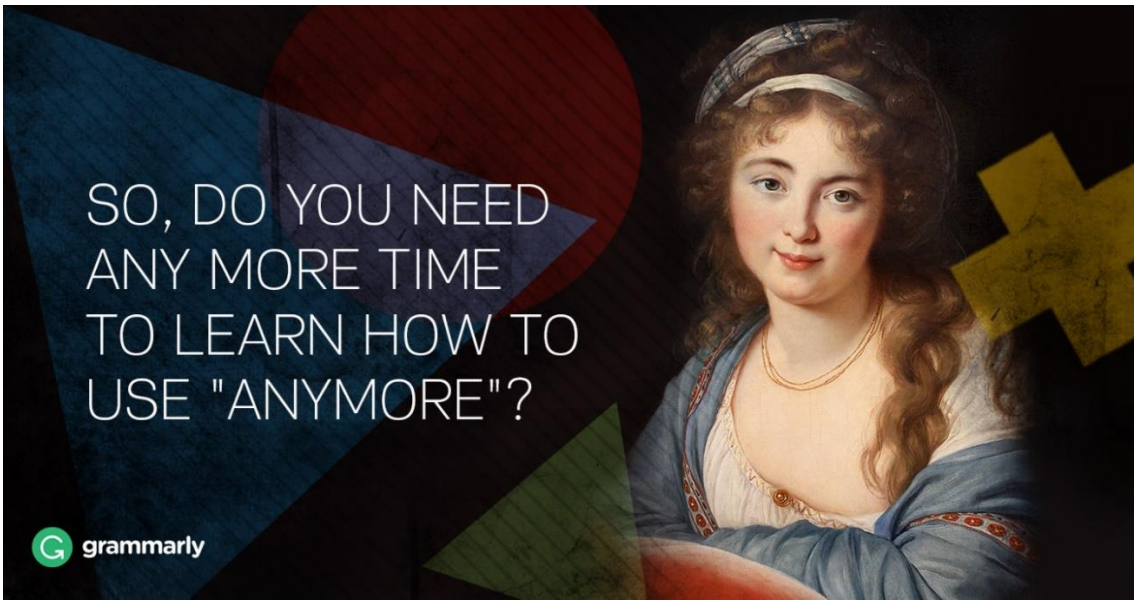

Figure 1: Example of visual content marketing used by Grammarly, an online proofreading and editing technology company. It challenges individuals who write in English to perfect their writing by using their platform.
(https://www.grammarly.com/blog/anymore-vs-any-more/)

When done correctly, marketing with visual content may generate three times as many leads and provide three times the return on investment for your business as sponsored search [Dayal, 2022]. It is a kind of communication that does not involve the use of words and is more successful when used with humans. In visual marketing, the primary focus is on the image or item that conveys the message, rather than the words themselves [Manic, 2015]. Images and other visual elements are becoming increasingly significant in the realm of digital marketing because they make a strong and immediate impression, stick with people for longer periods of time, and are easier to digest than other forms of material such as words [Dayal, 2022]. In addition, people are more likely to remember images and other visual elements than they are to remember words.

Studies have shown that our brains are able to process visual information up to 60,000 times faster than they can text and words, making it significantly more effective in the process [Sibley, 2017; Sadler, 2022]. When we read anything, we only remember about 20 percent of it, but when we see something, we recall up to 80 percent of it [Sadler, 2022]. Recent studies conducted at MIT have shown that the human brain is capable of processing and recognizing images in as little as 13 milliseconds [Sadler, 2022]. On the other hand, written material is processed in a way that is far slower and more linear. People living in today's world are said to be overloaded with information and have short attention spans as a result. As a consequence of this, a picture that is pleasing to the eye draws people into the subject matter and increases engagement.

Why does this hold true? In an ideal world, ninety percent of the information that your brain receives is visual, and around sixty-five percent of people are visual learners [Read, 2021; Sadler, 2022]. When compared to content that does not include a picture, the number of times that online content that includes relevant photos is viewed is nearly twice as high (94 percent higher) [Sadler, 2022]. In a similar vein, tweets that include images have a chance of being retweeted which is 150 percent higher than tweets that do not include images. Images account for 63 percent of all content shared on social media, and 54 percent of all Internet users have submitted an original photo or video at some point [Sadler, 2022].

Further, people are only likely to remember 10 percent of the information they are presented with three days after first hearing it [Medina, 2012]. They are able to remember 65 percent of the information three days later when it is presented to them in the form of a relevant image in addition to the same knowledge [Medina, 2012]. In addition, 46.1 percent of individuals believe that the design of a company's website is the most important factor to consider when determining the legitimacy of the business [Sibley, 2017]. This suggests that the utilization of photos is the single most important factor in producing useful material for social media. Because Google, the search engine giant, is aware of the huge power that can be derived from visual material, it is progressively incorporating graphics into the presentation of search results.

Within the realm of marketing, visual material holds the position of king. Visual marketing has many benefits, one of which is the immediate rise in brand recognition it provides. Studies have shown that it takes approximately fifty milliseconds to form an initial opinion of someone [Guest Author, 2021]. Marketers have, at long last, come to the understanding that it is no longer sufficient to merely focus on the content of one's communication; rather, they must also pay attention to the manner in which the content is communicated.

Seventy percent of companies' marketing budgets are currently allocated to content marketing, which may also incorporate graphic marketing tactics [Hubspot, 2022]. The toughest difficulties for 23.7 percent of content marketers are design and visual content [Enfroy, 2022]. While 49 percent and 22 percent of marketers regard visual marketing to be highly important and important, respectively, 19 percent of marketers confess that their strategy is meaningless without visual material (Mawhinney, 2022). Forty percent of marketers are certain that visual content will be heavily utilized by 51 percent to 80 percent of companies in the not-too-distant future [Khoja, 2022].

## 2.1 The Future of Visual Marketing

Your message cannot be adequately communicated through the use of words alone, nor can it be done with the use of an image alone. When you are able to connect a particular image to a certain message, you have the ability to really get into the heads of your audience and leave an imprint that will remain [Taylor, 2022]. One of the primary reasons why the world of marketing is ready for visual strategies is due to the fact that consumer technology places a strong emphasis on aesthetically pleasing components [Taylor, 2022]. Whenever we are looking at information, regardless of whether or not we are aware of it, our eyes are going to be drawn to the visual picture first. The longer we stare at something, the greater the likelihood that we will click over to the next page. As a consequence of this, the images you create need to be designed with a goal in mind and need to be made to stand out.

The future of marketing will require greater levels of personalization, the ability to operate in real-time, an increased reliance on technical applications, and the use of data to back up decisions. Marketers can no longer rely solely on the traditional marketing mix and marketing that is supported by data. When you include art, design, science, and technology into your marketing techniques, you can maximize the effectiveness of those strategies, such as attracting awareness, while also maintaining a fashionable and up-to-date image and making use of the technologies in a practical way [Lau, 2022].

Artificial intelligence, with the assistance of GANs (Generative adversarial networks), may effortlessly transform human thoughts into visual representations [Hughes, Zhu, & Bednarz, 2021]. Not only are AI-generated images causing a revolution in the art market, but they are also providing spectators with a more personalized experience [Lau, 2022]. The media of today is more complicated than it has ever been, and it advances at a rate that is quicker than it has ever been. No longer can marketing be done in isolation from technological advancements [Lau, 2022]. The time has come for marketers to fully embrace emerging technologies and close the gap that separates them from more traditional forms of marketing.

## 3 ARTIFICIAL INTELLIGENCE AND ART

### 3.1 Defining AI

The term "artificial intelligence" or "AI" refers to a powerful tool that is already reshaping every industry and field by compelling us to rethink how we integrate information, analyze data, and apply the insights gained to our decision-making. It is a foundational technology that can help modern businesses become more productive and competitive [Jin et al., 2021]. The average person is completely unfamiliar with the concept of artificial intelligence. One survey found that only 17 percent of 1,500 top-level American corporate leaders who were polled on artificial intelligence in 2017 claimed to have any prior knowledge of the subject matter [Davenport, Loucks, & Schatsky, 2017]. Each of the leaders had different levels of comprehension of what it was and how it would affect their own enterprises. Many of them were aware that artificial intelligence had the ability to completely transform how business is conducted, but they did not have a clear understanding of how the technology could be implemented within their own companies. Despite the fact that most people aren't familiar with it, artificial intelligence is a technology that is having far-reaching implications.

That said, there is no stand-alone definition of AI. Nonetheless, some researchers and IT experts have provided some definitions to help shed light on what artificial intelligence is and what it entails. According to Shubhendu and Vijay [2013], artificial intelligence refers to "machines that respond to stimulation consistent with traditional responses from humans, given the human capacity for contemplation, judgment, and intention." Ideally, these software systems make decisions that ordinarily demand a human degree of skill, and they assist individuals in either avoiding difficulties in the future or resolving them when they arise. Machines are intelligent, deliberate, and adaptable, just like human beings.

### 3.2 Defining Art

Just like AI, there is no one particular definition of art. It is a word whose definition has been sought after for ages, with the likes of Plato and Aristotle being the earliest contributors to the definition of the word. When we create works of art, we are not only expressing our innermost thoughts, feelings, intuitions, and desires, but also our unique perspectives on the world. It's a way of expressing deeply personal ideas that are difficult, if not impossible, to express in words. Because words alone cannot convey our meaning, we must seek out alternative means of expression. However, the art is not the content that we inject into our preferred mediums. The expression of ideas through any medium is the artistic expression.

There are also several other theories presented by several philosophers. There is a notion put forth by Croce [1904] that art is a superior intuition and, consequently, a superior form of knowledge attained through expression. Altieri [1987] further postulates that artistic expression is a creative and profound expression of human agency. Benjamin [2008] also believes that art is characterized by its aura, which is related to its singularity, but also to the audience's ability to share in that singularity through their own subjectivity. Danto [1964] believes that art is an intangible element, analogous to the "atmosphere of artistic theory." Dickie [1969] proposes a more pragmatic approach by stating that in order for art to exist, it must 1) be an artifact, and 2) have the status of "candidate for appreciation" bestowed upon it by some society or sub-group." Although these theories shed light on some of the characteristics shared by Art across cultures and time, they do not provide a complete or even comprehensive definition.

### 3.3 Combination of AI and Art

Artificial intelligence has advanced rapidly in recent years, with innumerable new examples of its uses emerging. The use of artificial intelligence in picture development is a useful stepping stone toward the introduction of artificial intelligence into the world of design and art. Ian Goodfellow and colleagues proposed Generative Adversarial Networks (GANs) in 2014 [Ragot & Martin, 2020]. These networks can produce better results by utilizing mutual game learning between the discriminator and the generator. Furthermore, when used for image synthesis, semantic segmentation, and other tasks, these networks perform better. In light of this possibility, cutting-edge scientists have experimented with applying artificial intelligence to the domains of art and design. For example, Qian Dong, Chen Yingqiao, and Huang Si, along with other graduate students from Zhejiang University's Software School's research and development team, created Xiang Ding —an artificial intelligence cultural and creative product [Han, 2021].

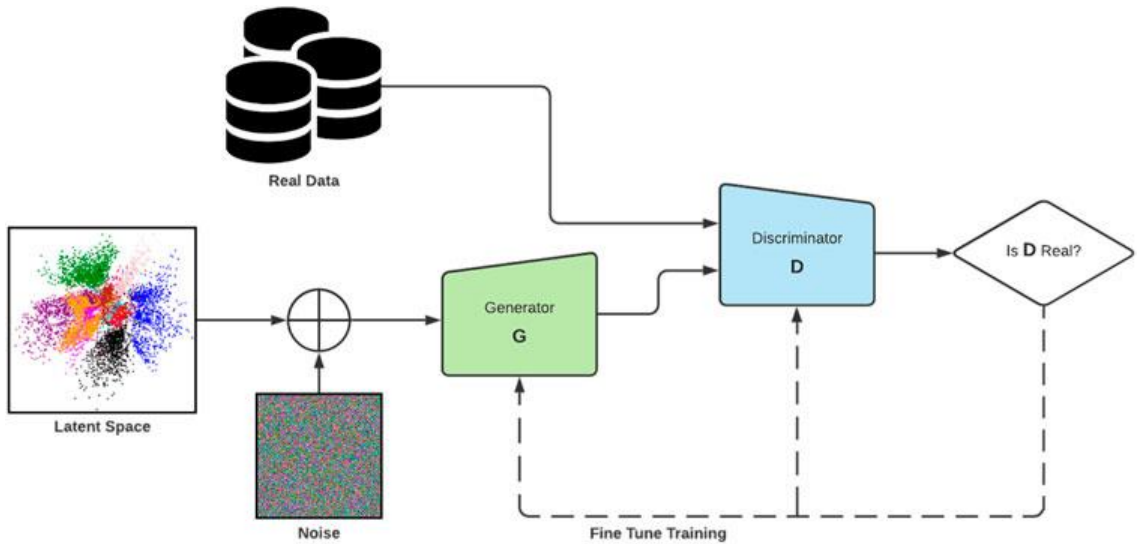

Figure 2: The original GAN architecture as described by Goodfellow et al. [2014].

The platform, which is based on the most recent artificial intelligence technology, provides a comprehensive process solution of personalized design, customization, and sales support for customers, enterprises in the cultural and tourism industry, and designers [Han, 2021]. Until now, the team has created cultural and tourism industry styles such as traditional culture styles, festival styles, world-famous painting styles, and independent artist co-branding styles. The organization is dedicated to putting high-quality artwork into the public light and providing a forum for unregistered artists to convert the intellectual property. Users can use the customization option given by AI Postcard to merge any digital images with the artist style provided by the site. This is only one example of a service. According to the calculations, the system is capable of instantly producing a completely new image in the artist's painting style.

Technology is an important part of the creative process of the art design. Digital technology, for example, acts as the foundation for digital media art. Digital technology largely uses modern computer and network information technology to edit and analyze optical and sound signals before converting these signals to digital information for archiving and administration. The fundamental building blocks of fashion design are pattern, structure, color, fabric, and so on. These design links should be placed in the strictly technical category. As a result, the term "art design" has acquired a scientific sense. A growing number of designers all across the world are studying a new AI-based design approach. This is because computer technology has advanced in the age of artificial intelligence and questions are now arising regarding computational creativity with regard to humans.

### 3.4 AI In Design

The use of artificial intelligence enables you to take care of a few routine tasks, which can save you work, time, and effort. It has the potential to do wonders for one's imagination and to assist in the generation of amazing thoughts in relation to design. Artificial intelligence is one of the low-cost approaches for growth with competence and swiftness, which means that it adds to refining speed and delivers superior design recommendations [Carton, 2022]. Artificial intelligence is the perfect kind of technology to assist designers create designs that are better and more innovative for the people who will be using them. The application of artificial intelligence in the design process can result in increased productivity and professionalism [Daniele & Song, 2019]. This technology allows designers to be more creative, and it also makes it easier to include interesting elements in their projects. It has the potential to save you time and effort by presenting you with fresh concepts for your project.

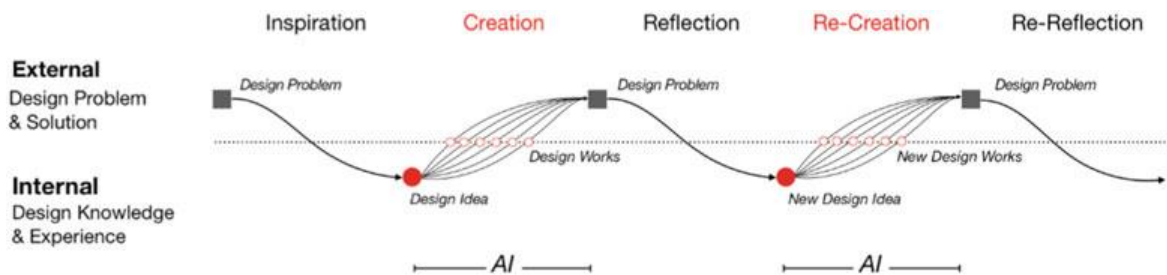

Figure 3: The AI augment creative design cycle [Zeng et al., 2019]. It shows how designers can make use of AI to supplement their design process.

The capacity for original thought that exists in human beings sets them apart from robots in the greatest way. The use of artificial intelligence helps bridge this gap [Mazzone & Elgammal, 2019]. In the recent past, artificial intelligence has emerged as the undisputed victor in the battle to understand user preferences and behaviors. It gives designers a road map to follow, which enables them to create designs that are consistent, user-friendly, and effective [Chen et al., 2019]. After identifying a pattern, designers can more easily generate various variants with the assistance of artificial intelligence. The algorithm first separates the individual colors and patterns that make up a design, and then it generates millions of different iterations using those colors and patterns in various combinations [Liu, Ren, & Liu, 2021]. Again, when they are at a loss for creative ideas, designers can find assistance from it. The abundant aesthetic options that are accessible might make the decision-making process simpler for both clients and designers. Additionally, it paves the way for creativity and enables designers to experiment with various color combinations and pattern arrangements.

According to Ahmed [2022] artificial intelligence will generally;
- Inspire creativity
- Help determine user preferences
- Help generate multiple variants
- Bring humanity to user experiences
- Personalize user interactions
- Assists people without basic or expert design skills

## 3.5 Generative Art

The process of algorithmically producing new ideas, forms, shapes, colors, or patterns is what we mean when we talk about generative art [Benney & Kistler, 2021]. The process of creativity begins with the formulation of rules that serve as parameters for the work. After that, a computer will generate new works in accordance with those rules on your behalf. Artists are making the most of their opportunities to be creative by utilizing computational tools to rapidly investigate, optimize, and test various creative design ideas. The process of design creates a balance between what is expected and what is unforeseen as well as between control and relinquishing that control. In spite of the fact that the procedures are determined, the outcomes cannot be predicted. The computer now possesses the ability to take humans by surprise. According to research conducted and estimations provided by professor and researcher Margaret Boden, "95 percent of what professional artists and scientists undertake is exploratory" [Benney & Kistler, 2021]. The use of generative systems is enabling researchers to cover far more territory in a shorter amount of time than in the past.

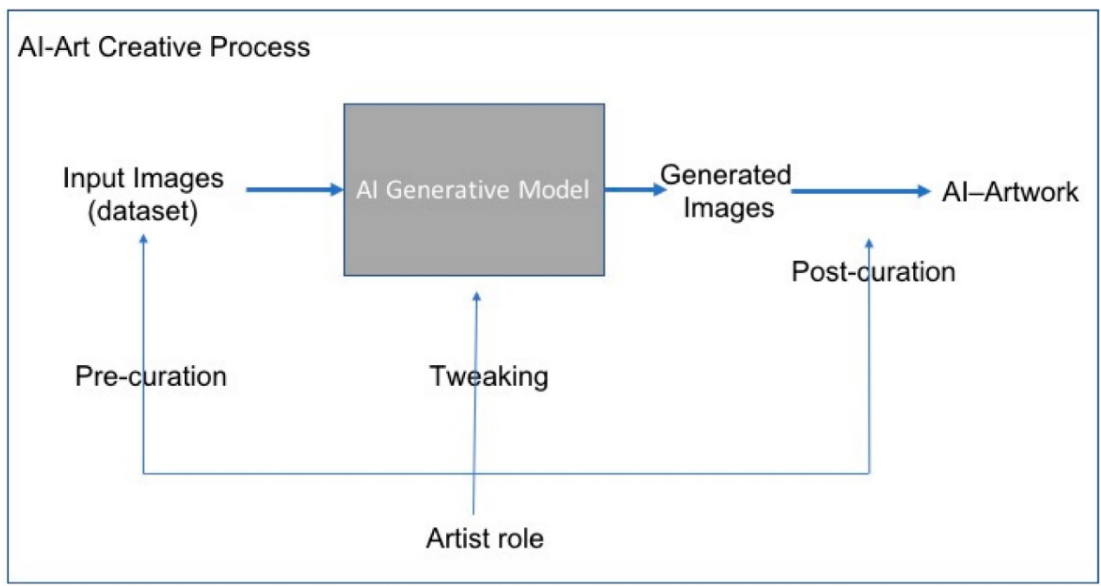

Figure 4: A block diagram showing the artist's role using the AI generative model in making art [Mazzone & Elgammal, 2019].

Ideally, generative art is primarily the yielding of control by the artist to an independent system [Benney & Kistler, 2021]. Programmable recipes known as evolutionary algorithms are used to create works of art by simulating the process of evolution. This involves "breeding" new concepts and "giving birth" to new shapes throughout many generations. An algorithm can be thought of as a recipe. The solution to any difficulty may be broken down into manageable steps. In addition, it is a crucial component in the overall structure of the field of computer science. Computers can compute algorithms at higher speeds and greater scales than the human brain, which enables artists to push the bounds of expression in whole new ways. Examples of this include generative fractal art as well as new immersive geometries that respond to sound and motion [Benney & Kistler, 2021]. Because this type of artwork is generated by a computer by following a predefined series of steps that are written in code, it is also frequently referred to as procedural art or code art.

The task of comprehending and producing meaningful procedural art is one that requires a significant amount of creative effort. Tools for generative creation have already been put to use in the production of a wide variety of things, including but not limited to: articles of journalistic news, paintings, music, poems, song lyrics, furniture, image, and video effects, industrial design, comics, illustrations, and even architecture. The accessibility of a wide variety of creative skills is being improved by the use of assistive creation technologies. People who do not currently possess any particular creative skills will acquire those skills, which will give rookie artists more power and decrease the barrier to entry for new work. Systems of generative creation are illuminating whole novel shapes and concepts that we never before could have conceived about. By changing the focus of artistic endeavors away from producing an end result and more toward the creation of a process, they are making available a whole new world of possibilities. They will motivate a new generation of artists to use tools to become more creative and more efficient with their time by utilizing computers to do the labor and develop near-unlimited variations in a given solution space. This will inspire a new generation of artists to use these tools.

Artificially intelligent tools will learn from us and become partners. They will offer ideas in real time, helping us sustain a flow state and reducing the likelihood of experiencing writer's block. Instead of being a straightforward replacement for human creativity, they will serve as a catalyst for it. Evolutionary algorithms are making it feasible to encode the fundamental processes of nature into the creation of art. This enables creatives to tap into the knowledge of the ages and manifest it in invigorating new ways, which is made possible by the evolution of the creative process.

**3.6 Generative Media**

It is just not sufficient to create content on a daily basis in this day and age, especially in a world where it appears like everyone is providing material. You need to develop content that is visually appealing, original, and high in volume, and you need to do all of this without exceeding your marketing budget. Users of Facebook upload 240,000 images every minute, while users of Instagram post around 65,000 photos each and every minute [Domo, 2021]. That's a huge amount of new content being added to the internet every single day, which could present a challenge for those who work in digital marketing. What is the answer? The limitless potential of generative media.

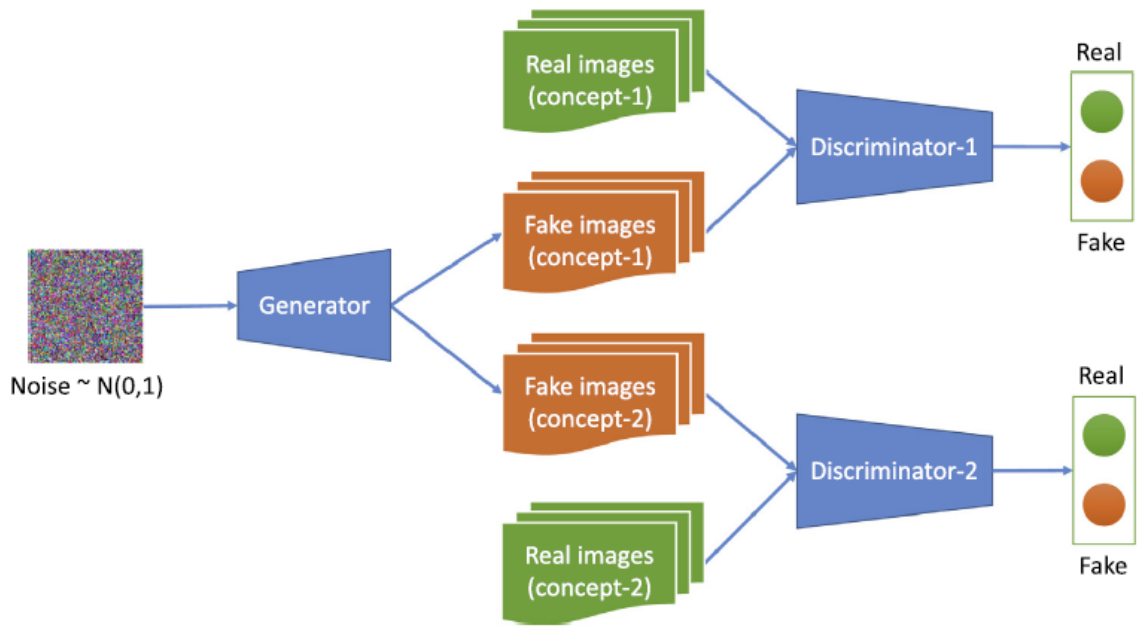

Figure 5: Generative adversarial networks model [L. Chen et al. / J. Vis. Commun. Image R. 61, 2019].

Code art, generative design, generative art, algorithm art, evolutionary art, synthetic art, and procedural art are all different names for the same thing: computer-generated artwork or images created by artificial intelligence and referred to collectively as "generative media" [Patel, 2021]. The artist gives the machine instructions to generate content based on a set of parameters or algorithms and the machine performs it. It's a joint effort by human artists and technological devices. By the year 2025, it is anticipated that the global market for generative design will reach 397.49 million USD [Fior Markets, 2019]. Having said that, this is not merely a visual trick. The term "generative media" refers to a set of practices that can be applied successfully in a variety of contexts and formats. Generative media can be utilized for the production of music streams, slogans, conversations, and videos for marketing purposes. In order to successfully improve one's marketing efforts, one can also successfully apply strategies that span many media.

In this sense, "content production" does not refer solely to the artistic expression of ideas when discussing generative media. It's all about providing a better visual experience for the user overall. Because of the immersive experience that is provided by the method, your levels of engagement may increase if you employ generative media. One very good example can be found in marketing. Seventy-one percent of individuals favor tailored advertisements [Kirkpatrick, 2016]. Additionally, according to a survey by HubSpot Research, over three-quarters of customers, or 72 percent, prefer watching videos to reading text when it comes to acquiring information about branded marketing [Hayes, 2022]. This indicates that the use of visual media has to be at the forefront of your marketing plan. Users can have a profoundly immersive advertising experience by mixing generative media with other techniques such as machine learning. This allows for the personalization of anything from emails to videos with unique names and preferences. You may produce the number and variety of content you need with the assistance of artificial intelligence or generative art, without having to sacrifice the quality of the output.

**3.7 Computational Creativity**

The study of both artificial and human creativity is a general area of artificial intelligence (AI) known as computational creativity [Ragot & Martin, 2020]. Can a computer become as creative as a human or more creative than a human being? To respond to this question, we need to first understand what creativity is. There is no one agreed-upon definition of creativity, especially within the context of machines, and there is little research to the same effect [Sbai et al., 2018]. Csikszentmihalyi [1997] defines creativity as a process that leads to the creation of an original and valued artifact that contributes to the advancement of culture. He goes on to deconstruct this notion by stating that creativity is a system composed of three components: a domain, which can be thought of as a collection of symbols and rules; a field, which can be thought of as a group of people with the expertise to evaluate given artifacts; and the individual who is the source of creativity. As a result, creativity can only develop as a result of the interaction between these three components.

Boden [2009] expands on Csikszentmihalyi's definition by introducing new terminology. She introduced the concept of P-creativity, which is novel to the individual who created the original artifact, and H-creativity, a P-Creative artifact that has never been seen before in history. She accomplished this by splitting the concept of novelty into psychological and historical meanings. Boden further distinguishes three types of creative thinking: combinatorial, transformative, and exploratory. These three types of creativity differ in terms of whether the search space is maintained in a static or fluid condition throughout the creation of objects [Barreto, Cardoso, & Roque, 2014].

Boden's previously established explanation for creativity was formalized by Wiggins [2006]. He starts by introducing a mathematical tuple that defines the universe of conceptions as well as the rules for exploring that universe. This tuple also offers guidelines for exploring the universe. It offers the groundwork for a framework for search algorithms that generalizes on traditional artificial intelligence searching methods. Wiggins even suggests that this framework is capable of performing more complex searches.

So, can computers be creative? Leonel Moura argues that computers have "some degree of creativity", whereas Harold Cohen does not ascribe to this [Audry & Ippolito, 2019]. Going by the above definitions, they cannot. Computers are believed to lack creativity because they lack human traits such as emotion. In fact, Colton et al. [2013] argue that human creativity and computer creativity should be kept separate. Perhaps we ought to look at the issue differently altogether by focusing more on the whole process as opposed to focusing on the end product [Oppenlaender, 2022]. Nonetheless, there is an ongoing corpus of research that studies how artificial creativity might be used to boost human creativity [Barreto, Cardoso, & Roque, 2014].

### 3.8 Can Machines Become Artists?

The progress made in Artificial Intelligence is achieving new heights of brilliance at an unparalleled rate. The use of these technologies in industries such as medicine, transportation, agriculture, retail, and security, to name a few, are subtly impacting the world. As a tool or a co-creative agent in tasks that have been the privilege of humans up until now, artificial intelligence is also assuming a growing role in the creative business as aforementioned. Throughout the past few years, we have witnessed the development of computational models for parametric architecture, generative fashion design, and procedural videogames, all of which have enabled designers to expand the scope of their creative potential.

Computers have already proven to be capable of creating something beyond their creator's imaginations [McCormack et al., 2014]. In point of fact, artists from all over the world have begun experimenting with open-source computational architectures such as Deep Neural Networks (DNN). This has resulted in the creation of artwork that is typically referred to as "Neural Art" or "AI Art." An artwork that was created by the computational artist Memo Akten using the program "Deepdream" was purchased at a benefit auction in February 2016 for the amount of 8,000 USD [Mordvintsev, Olah, and Tyka, 2015]. After only two years, the storyline of Neural Art is already moving in the direction of a science fiction setting.

The creative studio named Obvious developed an artwork that was wholly made through an algorithm. It was expected that the work would sell at Christie's for an amount ranging between 7,000 and 10,000 USD; instead, it sold for the astounding amount of 432,500 US dollars [Christie's, 2018]. This got the experts wondering whether or not we are witnessing the creation of a new art market. The fact that the members of the collective known as "Obvious," who are responsible for this work, do not have a background in art and do not appear as the artists but rather as the publishers, is an interesting fact. To a considerable extent, this point was utilized to speculate about the autonomy and the creative agency of the algorithm, which, for the general public, has now become the "true artist."

In 2021, Guo et al. introduced an intelligent system called Vinci. Vinci was primarily designed to help with the creation of advertisement posters. When it was subject to the Turing test, it was found that Vinci could make adverting posters that were no different from those made by human designers. Arguably, this was yet another proof that machines are slowly but gradually becoming artists in their own right.

The fact that the art market is showing interest in these pieces is a socioeconomic signal that raises more profound concerns about the relationship between art, technology, and society than can be answered by the bare facts of the market. Because of this incident, we have the opportunity to reconsider the roles that art and technology play in our lives, as well as the ways in which autonomous systems have the potential to transform the paradigm. Art is not a quantifiable element. Artists will always be social constructions so long as people attribute intentions to their art. In other words, we started with the wrong question when we asked if machines could become artists. In its place, we need to consider what kind of place artists, both real and imagined, human and silicon, and the viewers who envision them have in a world where more and more art is being produced by machines.

**3.9 Is AI Art Real Art?**

As more and more jobs are given to machines, the question arises over the outcomes obtained. People have attempted to develop machines that generate art (like in the case of Christie's above), and many works that were created by machines are already presented as works of art [Adajian, 2012]. Is it actually art, and if so, how does it compare to the works created by humans? Where exactly do these works and these creative acts stand in terms of their legal standing?

If someone claims that a computer can write music, can it be considered art in and of itself? What should we make of the assertions that, for example, a robot is able to draw pictures of people? We can observe the outcome as well as the performance, and there is a possibility that we see something that resembles art. But would you call it art? Is it possible, for example, that the robot is drawing? Concerning the legal standing of these creative processes and works of art, there is a degree of ambiguity.

It is interesting to note that it is not sufficient to respond to these questions by claiming that the outcomes of these creative and scientific experiments are merely "planned" and thus are not artistic [Coeckelbergh, 2016]. However, it is not as straightforward as that suggests. They are programmed in the sense that the algorithm, the code, is programmed; nonetheless, the end product—what is claimed to be the work of art—is not directly generated by a human being [Coeckelbergh, 2016]. This is because the algorithm and the code are both programmed. In this case, the artist is actually the algorithm itself, not the person.

The human is the creator of the code, not the work of art. The non-human creator is created by human creators, but the work that is created by the non-human agent is not directly created by humans. It would appear that the programmer is no longer the exclusive source of creativity, but rather that the technology has taken over this role. This is especially true in situations in which the machine possesses the ability to learn or when, in other respects, the process cannot simply be reduced to the execution of a written code. Arguably, it appears that machines are beginning to infiltrate a domain that was formerly reserved for people and that their outcome can actually be referred to as art.

**3.10 Do Humans Have a Preference for AI Art Over Human Art?**

Yes, they do. Several studies have been conducted where people have been shown artificial intelligence and human artworks and they have consistently been found to gravitate towards the artificial intelligence artworks. In a 2017 study, scientists confirmed that there was a perceived bias toward artworks made by machines as opposed to those made by humans [Ward, 2017]. Most of the participants said that they found the artificial intelligence pieces of art to be "novel, complex, and inspiring." Surprisingly, most could not even tell them apart. This goes to show just how good artificial intelligence art has become compared to human art.

**3.11 Will AI Replace Designers?**

There is a constant threat that artificial intelligence will completely replace human beings in the workplace. This effect is bound to be first felt in professions where you are required to follow regulations without using your imagination, such as engineering, law, and accounting, where machines would be able to quickly come up with 20 or 100 different answers to a given issue [Chowdhury, 2022]. This would be followed by all of the professions where humans are able to use their imaginations. As a direct result of this, there will be a significantly reduced demand for human resources. However, things are a little more complicated when it comes to the design industry.

Probability "guessing" is something that computers are really good at doing. Nevertheless, they are unable to contextualize designs and fail to evoke empathy from customers [Coeckelbergh, 2017]. It will be possible for designers to proceed through the production phases at a faster pace if they automate activities that involve repetitive actions, discover ways to free up more time for comprehension, empathy and focus more on the strategic aspects of the design process issue [Chowdhury, 2022]. As a consequence of this, projects will become more relevant, and experiences will become more tailored to the individual.

Vision, progression over time, and the ability to anticipate what will be required next are all essential aspects of design. Artificial intelligence will soon be able to replace designers, but it will be able to do so for the designers working today rather than the designers working in the future. On the positive side, it could be used as a creative partner and tool by designers to help them meet the constantly shifting demands of the industry [Sandry, 2016].

## 4 CONCLUSION

Both art and technology are, to varying degrees, reflections of who we are as individuals. On the one hand, the creation of art is a meta-language that allows us to express things that we are unable to articulate in any other way. It is a means by which we can make sense not only of the world that is all around us but also of ourselves as a species. In a similar vein, technology enables us to better understand how things work. It provides another lens through which to study human nature. In point of fact, the very concept of artificial intelligence is likely the most humanized form of technological advancement. Utilizing the power of artificial intelligence to build interfaces and systems that add to the creative toolkit of design practitioners is still in its infancy stage. However, there is a lot of potential for growth in this area. Even though work in this field is just getting started, several really useful tools are already beginning to take shape. In terms of the areas that researchers are concentrating their efforts on, trends are beginning to emerge. These areas include sketch-based interfaces, in situ design, and end-user–driven interface design.

Artificial intelligence will be a revolutionary tool in the design and marketing industry. It will help in the generation of user-based content and the automation of processes. It will also be a major driver in the creation of design tools and software, a move that will see the industry transform into a fast, efficient, and effective delivery service field. Looking at the already promising future of generative media and generative art, the future is only bound to get better. Visual marketing is about to become even more effective with the integration of artificial intelligence, helping convert a significantly higher larger of potential customers into loyal customers. However, this will happen if and only when artificial intelligence is viewed as a complement to human resources as opposed to a replacement.

The artificial intelligence models of today are still highly dependent on the input of humans, and the myth of complete creative autonomy does not appear to be imminent, at least according to certain experts in artificial intelligence. Creativity and imagination are defining features of our species, and we believe it is essential for artists to continue investigating technology tools such as artificial intelligence for the benefit of both the scientific community and the artistic community. However, if we want to automate creativity, expressivity, and imagination, we need to exercise extra caution and have a conversation that includes both the sciences and the humanities. The complexity of this subject ought to expand beyond the technological community and reach into the bigger fields of business, philosophy, neurology, and the arts.